\documentclass[11pt]{article}

\usepackage{fullpage}
\usepackage[ruled, linesnumbered, vlined, commentsnumbered]{algorithm2e}
\usepackage{graphicx,subfig} % figure related
\usepackage{amsfonts,amssymb,amsmath,amsthm,amsopn}	% math related
\usepackage{hhline,booktabs,colortbl,multirow,tabularx,threeparttable} % table related
\usepackage[listings,skins,breakable]{tcolorbox}
\usepackage[shortlabels]{enumitem}

\usepackage{authblk}
\usepackage{footnote}
\usepackage{hyperref}
\usepackage{prettyref}
\usepackage{cite}
\usepackage{setspace}
\usepackage{color}
\usepackage{xcolor}  % Required for custom colors
\usepackage{scrpage2}
\usepackage{geometry}

\usepackage{tikz}
\usepackage{pgfplots}
\usetikzlibrary{positioning,shapes,shadows,arrows,calc}
\tikzstyle{component}=[rectangle, draw=black, rounded corners, fill=blue!40, drop shadow, text centered, anchor=north, text=white, minimum height=1cm]
\tikzstyle{arrow}=[->, thick]

\pgfplotsset{compat=1.12}
\usetikzlibrary{intersections}
\usetikzlibrary{pgfplots.statistics}
\usepgfplotslibrary{fillbetween}

\def\hlinew#1{%
  \noalign{\ifnum0=`}\fi\hrule \@height #1 \futurelet
   \reserved@a\@xhline}
\makeatother

\definecolor{myblue}{RGB}{34,31,217}
\definecolor{mycyan}{gray}{.7}
\definecolor{Gray}{gray}{0.9}

\newcommand{\pref}{\prettyref}

\newrefformat{fig}{Fig.~\ref{#1}}
\newrefformat{tab}{Table~\ref{#1}}
\newrefformat{sec}{Section~\ref{#1}}
\newrefformat{app}{Appendix~\ref{#1}}
\newrefformat{alg}{Algorithm~\ref{#1}}
\newrefformat{property}{Property~\ref{#1}}
\newrefformat{theorem}{Theorem~\ref{#1}}
\newrefformat{corollary}{Corollary~\ref{#1}}
\newrefformat{proposition}{Proposition~\ref{#1}}
\newrefformat{def}{Definition~\ref{#1}}
\newrefformat{eq}{equation~(\ref{#1})}

\title{\vspace{-1ex}\LARGE\textbf{DeepSQLi: Deep Semantic Learning for Testing SQL Injection}\footnote{This manuscript is accepted for publication in ISSTA 2020. The copyright of this paper has been permanently transferred to ACM. Li is the corresponding author of this paper. All authors made commensurate contributions to this paper. Li designed and supervised the research. Liu built the system and carried out experiments. Li and Chen interpreted data and wrote the manuscript.}}

\author[$\dag$]{\normalsize Muyang Liu}
\author[$\ddag$]{\normalsize Ke Li}
\author[$\ast$]{\normalsize Tao Chen}
\affil[$\dag$]{\normalsize College of Computer Science and Engineering, University of Electronic Science and Technology of China, 611731, Chengdu, China}
\affil[$\ddag$]{\normalsize Department of Computer Science, University of Exeter, EX4 4QF, Exeter, UK}
\affil[$\ast$]{\normalsize Department of Computer Science, Loughborough University, Loughborough, LE11 3TU, UK}
\affil[$\ast$]{\normalsize Email: \texttt{k.li@exeter.ac.uk}}
\date{}

\begin{document}
\maketitle

\vspace{-3ex}
{\normalsize\textbf{Abstract: }Security is unarguably the most serious concern for Web applications, to which SQL injection (SQLi) attack is one of the most devastating attacks. Automatically testing SQLi vulnerabilities is of ultimate importance, yet is unfortunately far from trivial to implement. This is because the existence of a huge, or potentially infinite, number of variants and semantic possibilities of SQL leading to SQLi attacks on various Web applications. In this paper, we propose a deep natural language processing based tool, dubbed \texttt{DeepSQLi}, to generate test cases for detecting SQLi vulnerabilities. Through adopting deep learning based neural language model and sequence of words prediction, \texttt{DeepSQLi} is equipped with the ability to learn the semantic knowledge embedded in SQLi attacks, allowing it to translate user inputs (or a test case) into a new test case, which is semantically related and potentially more sophisticated. Experiments are conducted to compare \texttt{DeepSQLi} with \texttt{SQLmap}, a state-of-the-art SQLi testing automation tool, on six real-world Web applications that are of different scales, characteristics and domains. Empirical results demonstrate the effectiveness and the remarkable superiority of \texttt{DeepSQLi} over \texttt{SQLmap}, such that more SQLi vulnerabilities can be identified by using a less number of test cases, whilst running much faster.}

{\normalsize\textbf{Keywords: } }Web security, SQL injection, test case generation, natural language processing, deep learning

% !TeX root=main.tex
\section{Introduction}
\label{sec:introduction}

Web applications have become increasingly ubiquitous and important since the ever prevalence of distributed computing paradigms, such as Cyber-Physical Systems and Internet-of-Things. Yet, they are unfortunately vulnerable to a variety of security threats, among which SQL injection (SQLi) has been widely recognised as one of the most devastating threats. Generally speaking, SQLi is an injection attack that embeds scripts in user inputs to execute malicious SQL statements over the relational database management system (RDBMS) running behind a Web application. As stated in the Akamai report\footnote{~\href{https://www.akamai.com/us/en/multimedia/documents/state-of-the-internet/soti-security-web-attacks-and-gaming-abuse-executive-summary-2019.pdf}{https://www.akamai.com/}}, SQLi attacks constituted 65.1\% of the cyber-attacks on Web applications during November 2017 to March 2019. It also shows that the number of different types of Web attacks (e.g., XSS, LFI and PHPi) has ever increased, but none of them have been growing as fast as SQLi attacks. Therefore, detecting and preventing SQLi vulnerabilities are of ultimate importance to improve the reliability and trustworthiness of modern Web applications.

There are two common approaches to protect Web applications from SQLi attacks. The first one is customised negative validation, also known as input validation. Its basic idea is to protect Web applications from attacks by forbidden patterns or keywords manually crafted by software engineers. Unfortunately, it is difficult, if not impossible, to enumerate a comprehensive set of validation rules that is able to cover all types of attacks. The second approach is prepared statement that allow embedding user inputs as parameters, also known as placeholders. By doing so, attackers are difficult to embed SQLi code in user inputs since they are treated as value for the parameter. However, as discussed in~\cite{Maor2004} and~\cite{Mcdonald02}, prepared statement is difficult to design given the sophistication of defensive coding guideline. In addition, there are many other terms, such as dynamic SQL of DDL statement (e.g., \texttt{create}, \texttt{drop} and \texttt{alter}) and table structure (e.g, names of columns, tables and schema) cannot be parameterised.
%A good way to avoid SQLi attacks is parameterization which allows embedding the user inputs as parameters (placeholders). In this way, user inputs are treated as value for the parameter. Attackers cannot inject SQL code in user inputs. However, there are still some people who don't use parameterization or don’t use it correctly. Moverover, dynamic SQL of DDL statements (such as create, drop and alter) and identifiers in SQL statement (such as names of columns, tables and schemas) can’t be parameterized. Therefore, the detection and prevention of SQLi on Web applications is of ultimate importance.
%In practice, an attacker can take benefit of the SQLi vulnerabilities to bypass the input validation \textemdash\ the key protection mechanism of the Web applications, enabling one to amend the data record, retrieve unauthorized data or even control the RDBMS. Therefore, the detection and prevention of SQLi on Web applications is of ultimate importance.

Test case generation, which build test suites for detecting errors of the system under test (SUT), is the most important and fundamental process of software testing. This is a widely used practice to detect SQLi vulnerabilities where test suites come up with a set of malicious user inputs that mimic various successful SQLi attacks, each of which forms a test case. However, enumerating a comprehensive set of semantically related test cases to fully test the SUT is extremely challenging, if not impossible. This is because there are a variety of SQLi attacks, many complex variants of which share similar SQL semantic. For example, the same attack can be diversified by changing the encoding form, which appears to be different but is semantically equivalent, in order to evade detection.

Just like human natural language, malicious SQL statements have their unique semantic. Therefore, test case generation for detecting SQLi vulnerabilities can take great advantages by exploiting the knowledge from such semantic naturalness. For example, given a Web form with two user input fields, i.e., \texttt{username} and \texttt{password}, the following SQL statement conforms to a SQLi attack:

\vspace{0.2em}
\noindent
\framebox{\parbox{\dimexpr\linewidth-2\fboxsep-2\fboxrule}{
\texttt{\itshape \textbf{{SELECT}} * \textbf{{FROM}} members \textbf{{WHERE}} username=\lq\underline{\textcolor{blue}{admin\rq+OR+\lq1\rq=\lq1}}\rq\ \textbf{{AND}} password=\lq \underline{\textcolor{blue}{\rq--}}\rq}
}}
\vspace{0.01em}

\noindent
where the underlined parts are input by a user and constitute a test case that leads to an attack to the SUT. Given this SQLi attack, we are able interpret some semantic knowledge as follows.
\begin{itemize}
    \item This is a tautology attack that is able to use any tautological clause, e.g., \textcolor{blue}{\texttt{OR 1=1}}, to alter the statement in a semantically equivalent and syntactically correct manner without compromising its maliciousness.
    \item To meet the SQL syntax, an injection needs to have an appropriate placement of single quotation to complete a SQL statement. Therefore, the attack should be written as \textcolor{blue}{\texttt{admin\rq+ OR+\lq1\rq=\lq1}}. In addition, the unnecessary part of the original statement can be commented by \textcolor{blue}{\texttt{--}}.
    \item In practice, due to the use of some input filters like firewalls, blank characters will highly likely be trimmed by modern Web applications thus leading to the failure of \textcolor{blue}{\texttt{admin\rq\ OR 1=1}} to form a tautology attack. By replacing those blank characters with \textcolor{blue}{\texttt{+}}, which is semantically equivalent, the attacker is able to disguise the attack in a more sophisticated manner.
\end{itemize}    
Although semantic knowledge can be interpreted by software engineers, it is far from trivial to leverage such knowledge to automate the test case generation process.

Traditional test case generation techniques mainly rely on software engineers to specify rules to create a set of semantically tailored test cases, either in a manual ~\cite{CurpheyA06,KiezunGJE09} or \textit{semi}-automatic manner~\cite{TianYXS12,BenikhlefWG16,AppeltNPB18}. Such process is of limited flexibility due to the restriction of human crafted rules. Furthermore, it is expensive in practice or even be computationally infeasible for modern complex Web applications.

%Although there have been some initiatives that provide a \textit{semi}-automatic way to generate test cases by incorporating domain-specific knowledge~\cite{TianYXS12,BenikhlefWG16,AppeltNPB18}, the so-called knowledge is usually hard coded by experienced engineers in terms of rules and syntax beforehand, which can limit the flexibility of the approach.

Recently, there has been a growing interest of applying machine learning algorithms to develop artificial intelligence (AI) tools that automate the test case generation process~\cite{SkaruzS07,KimL14,Sheykhkanloo17,DoshiAF18} and~\cite{LiXLRWCZYS18}. This type of methods requires limited human intervention and do not assume any fixed set of SQL syntax. However, they are mainly implemented as a classifier that is used to diagnose whether a SQL statement (or part of it) is a valid statement or a malicious injection. To the best of our knowledge, none of those existing AI based tools are able to proactively generate semantically related SQLi attacks during the testing phase. There have been some attempts that take semantic knowledge into consideration. For example, \cite{Sheykhkanloo17} developed a classifier that considers the semantic abnormality of the \textcolor{blue}{\texttt{OR}} phrase (e.g., \textcolor{blue}{\texttt{OR 1=1}} or \textcolor{blue}{\texttt{OR 'i' in ('g', 'i')}}) in a tautology attack. Unfortunately, this method ignores other alternatives, which might be important when semantically generating SQLi attacks, to create tautology (e.g., we can use \textcolor{blue}{\texttt{--}} to comment out other code)
.

Bearing the above considerations in mind, this paper proposes a deep natural language processing (NLP) based tool\footnote{~All source code and experiment data can be accessed at our anonymous repository:~\href{https://github.com/COLA-Laboratory/issta2020}{https://github.com/COLA-Laboratory/issta2020}}, dubbed \texttt{DeepSQLi}, which learns and exploits the semantic knowledge and naturalness of SQLi attacks, to automatically generate various semantically meaningful and maliciously effective test cases. Similar to the machine translation between dialects of the same language, \texttt{DeepSQLi} takes a set of normal user inputs or existing test case for a SQL statement (one dialect) and \textit{translates} it into another test case (another dialect), which is semantically related but potentially more sophisticated, to form a new SQLi attack.

\vspace{0.5em}
\noindent
\underline{\textbf{\textit{Contributions}}.} The major contributions of this paper are:
\begin{itemize}
    \item \texttt{DeepSQLi} is a fully automatic, end-to-end tool empowered by a tailored neural language model trained under the \texttt{Trans}-\texttt{former}~\cite{VaswaniSPUJGKP17} architecture. To the best of our knowledge, this work is the first of its kind to adopt \texttt{Transformer} to solve problems in the context of software testing.
    \item To facilitate the semantic knowledge learning from SQL statement, five mutation operators are developed to help enrich the training dataset. Unlike the classic machine translation where only the sentence with the most probable semantic match would be of interest, in \texttt{DeepSQLi}, we extend the neural language model with \texttt{Beam search}~\cite{RaychevVY14}, in order to generate more than one semantically related test case based on the given test case/normal inputs that needs translation. This helps to generate a much more diverse set of test cases, and thus providing larger chance to find more vulnerabilities.
    \item The effectiveness of \texttt{DeepSQLi} is validated on six real-world Web applications selected from various domains. They are with various scales and have a variety of characteristics. The results show that \texttt{DeepSQLi} is better than \texttt{SQLMap}, a state-of-the-art SQLi testing automation tool, in terms of the number of vulnerabilities detected and exploitation rate whilst running up to $6\times$ faster.
\end{itemize}

\noindent\underline{\textbf{\textit{Novelty}}.} What make \texttt{DeepSQLi} \textit{unique} are:
\begin{itemize}
    \item It is able to translate any normal user inputs into some malicious inputs, which constitute to a test case. Further, it is capable of translating an existing test cases into another semantically related, but potentially more sophisticated test case to form a new SQLi attack.
    \item If the generated test case cannot achieve a successful SQLi attack, it would be fed back to the neural language model as an input. By doing so, \texttt{DeepSQLi} is continually adapted to create more sophisticated test cases, thus improving the chance to find previously unknown and deeply hidden vulnerabilities.
\end{itemize}

The remaining paper is organised as follows. \pref{sec:NLM} provides a pragmatic tutorial of neural language model used for SQLi in this paper. \pref{sec:approach} delineates the implementation detail of our proposed \texttt{DeepSQLi}. \pref{sec:experiment} presents and discusses the empirical results on six real-world Web applications. \pref{sec:threats} exploits the threats to validity and related works are overviewed in~\pref{sec:survey}. \pref{sec:conclusion} summarises the contributions of this paper and provides some thoughts on future directions.

\section{Deep Natural Language Processing for SQLi}
\label{sec:NLM}

In this section, we elaborate on the concepts and algorithms that underpin \texttt{DeepSQLi} and discuss how they were tailored to the problem of translating user inputs into test cases.

\subsection{Neural Language Model for SQLi}
\label{subsec:NLmodel}

Given a sequence of user inputs $I_u=\{w_1,\dots,w_N\}$ where $w_i$ is the $i$-th token, a language model aims to estimate a joint conditional probability of tokens of $I_u$. Since a direct calculation of this multi-dimensional probability is far from trivial, it is usually approximated by $n$-gram models~\cite{BrownPPLM92} as:
\begin{equation}
    \begin{aligned} 
    P\left(w_{1},\cdots, w_{N}\right) &=\prod_{i=1}^N P\left(w_i | w_{1},\cdots,w_{i-1}\right)\\
    &\approx \prod_{i=1}^N P(w_i|w_{i-n+1},\cdots,w_{i-1})
    \end{aligned},
    \label{eq:ngram}
\end{equation}
where $N$ is the number of consecutive tokens. According to~\pref{eq:ngram}, we can see that the prediction made by the $n$-gram model is conditioned on its $n-1$ predecessor tokens. However, as discussed in~\cite{GuthrieA0GW06}, $n$-gram models are suffered from a sparsity issue where it does not work when predicting the first token in a sequence that has no predecessor.

To mitigate such issue, \texttt{DeepSQLi} makes use of neural language model as its fundamental building block. Generally speaking, it is a language model based on neural networks along with a probability distribution over sequences of tokens. In our context, such probability distribution indicates the likelihood to which a sequence of user inputs conform to a SQLi attack. For example, a sequence of inputs \textcolor{blue}{\texttt{admin\rq+OR+\lq1\rq=\lq1}} will have a higher probability since it is able to conform to a SQLi attack; whereas another sequence of inputs \textcolor{blue}{\texttt{OR SELECT AND 1}}, which is semantically invalid from the injection point of view, will have a lower probability to become a SQLi attack.

Comparing to the $n$-gram model, which merely depends on the probability, the neural language model represents the tokens of an input sequence in a vectorised format, as known as word embedding which is an integral part in neural language model training. Empowered by deep neural networks, a neural language model is more flexibility with predecessor tokens having longer distances thus is resilient to data sparsity. 

In \texttt{DeepSQLi}, we adopt a neural language model for token-level sequence modeling, given that a token is the most basic element in SQL syntax. In other words, given a sequence of user inputs $I_u$, the neural language model aims to estimate the joint probability of the inclusive vectorised tokens.  

\subsection{Multi-head Self-Attention in Neural Language Model}
\label{sec:attention}

Attention mechanisms, which allow modelling of dependencies without regarding to their distance in sequences, have recently become an integral part of sequence generation tasks~\cite{MnihHGK14}. Among them, self-attention is an attention mechanism that has the ability to represent relationship between tokens in a sequence. For example, we can better understand the token \textcolor{blue}{"1"} in the sequence \textcolor{blue}{"\texttt{OR 2 > 1}"} by answering three questions: "\textit{why to compare (i.e., what kind of attack)}", "\textit{how to compare}" and "\textit{who to compare with}". Self-attention have been successfully applied to deal with various NLP tasks, such as machine translation~\cite{VaswaniSPUJGKP17}, speech recognition~\cite{DongXX18} and music generation~\cite{HuangVUSHSDHDE19}. In \texttt{DeepSQLi}, the self-attention is calculated by the scaled dot-product attention proposed by Vaswani et al~\cite{VaswaniSPUJGKP17}:
%In particular, to improve NLP performance, word embedding is carried out as the first step before building a natural language model like self-attention. It maps each token of the input sequence to a vector thus vectors of word embedding become a distributed representation of tokens. 
\begin{equation}
\begin{aligned}
\mathbf{Q}=\mathbf{X}\mathbf{W}_Q, \mathbf{K}&=\mathbf{X}\mathbf{W}_K, \mathbf{V}=\mathbf{X}\mathbf{W}_V\\
A(\mathbf{Q}, \mathbf{K}, \mathbf{V})&=\operatorname{softmax}\left(\frac{\mathbf{Q} \mathbf{K}^T}{\sqrt{d}}\right) \mathbf{V}
\end{aligned},
\label{eq:selfatt}
\end{equation}
\noindent where $\mathbf{X}$ is the word embedding, i.e., the vector representation, of the input sequence, $\mathbf{Q}$ is a matrix consists of a set of packed queries, $\mathbf{K}$ and $\mathbf{V}$ are keys and values matrices whilst $d$ is the dimension of the key. In particular, $\mathbf{Q}$, $\mathbf{K}$ and $\mathbf{V}$ are obtained by multiplying $\mathbf{X}$ by three weight matrices $\mathbf{W}_Q$, $\mathbf{W}_K$ and $\mathbf{W}_V$. 

In order to learn more diverse representations, we apply a multi-head self-attention in \texttt{DeepSQLi} given that it has the ability to obtain more information from the input sequence by concatenating multiple independent self-attentions. Specifically, a multi-head self-attention can be formulated as:
\begin{equation}
 MA(\mathbf{Q}_X,\mathbf{K}_X,\mathbf{V}_X)=[A_1(\mathbf{Q}_1,\mathbf{K}_1,\mathbf{V}_1)\otimes\cdots\otimes A_h(\mathbf{Q}_h,\mathbf{K}_h,\mathbf{V}_h)]\cdot\mathbf{W}_a,
 \label{eq:multihead}
\end{equation}
\noindent
where $\otimes$ is a concatenation operation, $\mathbf{W}_a$ is a weight matrix and $h$ is the length of parallel attention layers, also known as heads. Since each head is a unique linear transformation of the input sequence representation as queries, keys and values, the concatenation of multiple independent heads enables the information extraction from different subspaces thus leading to more diverse representations.

\vspace{2mm}
\begin{figure*}[t!]
  \centering
  \includegraphics[width=\linewidth]{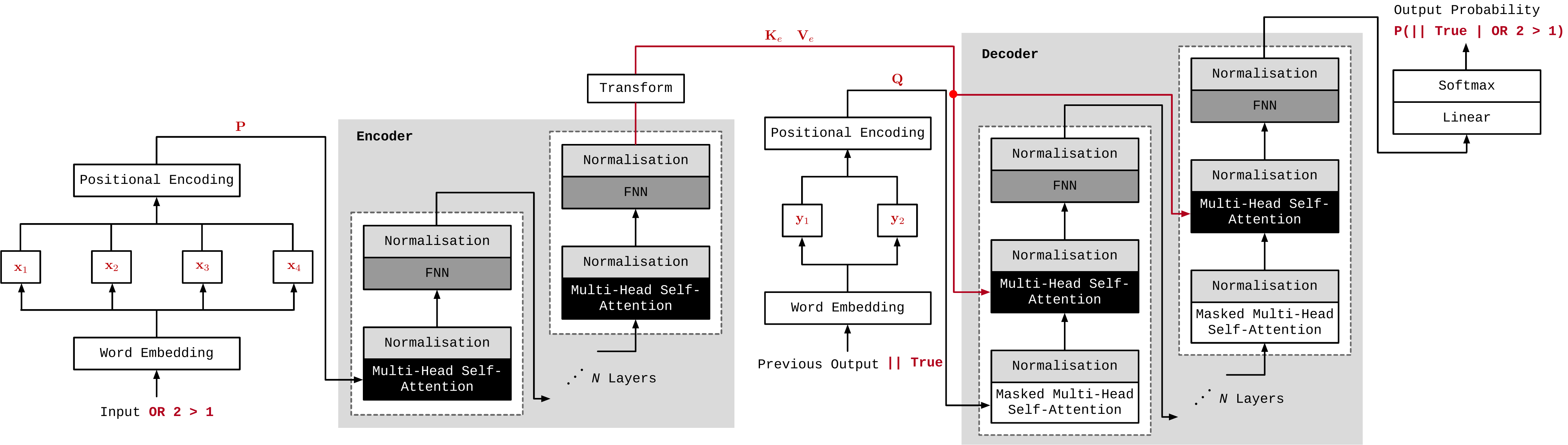}
  \caption {An illustrative working flowchart of the encoder-decoder (\texttt{Seq2Seq}) model in \texttt{DeepSQLi}.}
  \label{fig:NLM}
\end{figure*}

\subsection{Encoder-Decoder (Seq2Seq) model}
\label{subsec:seq2seq}

In order to train the neural language model, we adopt \texttt{Seq2Seq}\textemdash a general framework consists of an encoder and a decoder\textemdash in \texttt{DeepSQLi}. In particular, \texttt{Transformer}~\cite{VaswaniSPUJGKP17} is used to build the \texttt{Seq2Seq} model in \texttt{DeepSQLi} instead of those traditional recurrent neural network (RNN)~\cite{VinyalsKKPSH15} and convolutional neural network (CNN)~\cite{KalchbrennerGB14}, given its state-of-the-art performance reported in many \texttt{Seq2Seq} tasks.
% In particular, the \texttt{Seq2Seq} model used in this paper is built upon the \texttt{Transformer}~\cite{VaswaniSPUJGKP17} rather more traditional recurrent neural network (RNN)~\cite{VinyalsKKPSH15} or convolutional neural network (CNN)~\cite{KalchbrennerGB14}. \texttt{Transformer} model replaces RNN and CNN with multi-head self-attention to get rid of recurrence and convolutions, which has achieved state-of-the-art performance in Seq2Seq tasks~\cite{VaswaniSPUJGKP17}.
%\textcolor{red}{How do we decide on the number of FNN and the number of layers in encoder/decider?}

% In \texttt{DeepSQLi}, we adopt Seq2Seq\textemdash a general framework include an encoder and a decoder \textemdash to train our neural language model, which can then translate a sequence of normal user inputs (or source test case) into another semantically related sequence of test case that mimics SQLi attacks. 

\pref{fig:NLM} shows an illustrative flowchart of the \texttt{Seq2Seq} model in \texttt{DeepSQLi}. The encoder takes a sequence of vector representations of $N$ tokens, denoted as $\mathbf{X}=\{\mathbf{x}_1,\cdots,\mathbf{x}_N\}$, which embed semantic information of tokens; whilst the input of the decoder is another sequence of vector representations of $\overline{N}$ tokens, denoted as $\mathbf{Y}=\{\mathbf{y}_1,\cdots,\mathbf{y}_{\overline{N}}\}$. Note that $N$ is not necessary to be equal to $\overline{N}$. The purpose of this \texttt{Seq2Seq} model is to learn a conditional probability distribution over the output sequence conditioned on the input sequence, denoted as $P(\mathbf{y}_1,\cdots,\mathbf{y}_{\overline{N}}|\mathbf{x}_1,\cdots,\mathbf{x}_N)$. As for the example shown in~\pref{fig:NLM} where the input sequence is \textcolor{blue}{\texttt{OR 2 > 1}} and the output sequence is \textcolor{blue}{\texttt{|| True}}, the conditional probability is $P\left(\texttt{\textcolor{blue}{|| True}} | \texttt{\textcolor{blue}{OR 2 > 1}}\right)$.

More specifically, the encoder in the left hand side of~\pref{fig:NLM} consists of $N$ identical layers, each of which has a multi-head self-attention mechanism sub-layer and a deep feed-forward neural network (FFN) sub-layer. The first sub-layer is the multi-head self-attention. As show in~\pref{fig:NLM}, the output of the multi-head self-attention, calculated by~\pref{eq:multihead}, is denoted as $\mathbf{Z}_1$. It is supplemented with a residual connection $\varepsilon_{\mathbf{Z}_1}$ to come up with the output $\mathbf{N}_1$ after a layer-normalisation. This process can be formulated as:
\begin{equation}
\mathbf{N}_1=\texttt{layer-normalisation}(\mathbf{Z}_1+\varepsilon_{\mathbf{Z}_1}).
\end{equation}
Afterwards, $\mathbf{N}_1$ is fed to the second sub-layer, i.e., a FNN, to carry out a non-linear transformation. Specifically, the basic mechanism of the FNN is formulated as:
\begin{equation}
\mathbf{z}_2=\mathrm{FFN}(\mathbf{n}_1)=\max \left(0, \mathbf{n}_1 W_{1}+b_{1}\right) W_{2}+b_{2},
\end{equation}
\noindent where $\mathbf{n}_1$ is a vector of $\mathbf{N}_1$. Thereafter, the output of the FNN, i.e., $\mathbf{Z}_2$, will be transformed to two matrices $\mathbf{K}_e$ and $\mathbf{V}_e$ after being normalized to $\mathbf{N}_2$. It is worth noting that $\mathbf{K}_e$ and $\mathbf{V}_e$ are the output of the encoder whilst they embed all information of the input sequence \textcolor{blue}{"\texttt{OR 2 > 1}"}.

As for the decoder shown in the right hand side of~\pref{fig:NLM}, it takes $\mathbf{K}_e$ and $\mathbf{V}_e$ output from the encoder as a part of inputs for predicting a sequence of semantically translated vector representation $\mathbf{Y}$. The decoder is also composed of a stack of $N$ identical layers, each of which consists of a masked multi-head self-attention, a multi-head self-attention and a FNN sub-layers. In particular, the computational process of the multi-head self-attention and the FNN sub-layers is similar to that of the encoder, except that $\mathbf{K}_e$ and $\mathbf{V}_e$ are used as the $\mathbf{K}$ and $\mathbf{V}$ of \pref{eq:selfatt} in the multi-head self-attention sub-layer. As for the masked multi-head self-attention sub-layer, it is used to avoid looking into tokens after the one under prediction. For example, the multi-head self-attention masks the second token \texttt{\textcolor{blue}{"True"}} when predicting the first one \texttt{\textcolor{blue}{"||"}}.

In principle, the \texttt{Transformer} used to do \texttt{Seq2Seq} allows for significantly more parallel processing and has been reported as a new state-of-the-art. Unlike the RNN, which takes tokens in a sequential manner, the multi-head attention computes the output of each token independently, without considering the order of words. Since the SQLi inputs used in \texttt{DeepSQLi} are sequences with determined semantics and syntax, it may leads to a less effective modelling of the sequence information without taking any order of tokens into consideration. To take such information into account, the \texttt{Transformer} supplements each input embedding with a vector called positional encoding (PE). Specifically, PE is calculated by sine and cosine functions with various frequencies:
\begin{equation}
\begin{aligned}
\mathbf{PE}_i&=\left\{\begin{matrix}
\{\sin(\frac{i}{10000\frac{2}{d_e}}),\cdots,\sin(\frac{i}{10000\frac{2d_e}{d_e}})\},\quad \text{if}\: i\%2==0\\ 
\{\cos(\frac{i}{10000\frac{2}{d_e}},\cdots,\cos(\frac{i}{10000\frac{2d_e}{d_e}})\},\quad \text{if}\: i\%2==1
\end{matrix}\right.,
% \mathbf{x}_i&=\mathbf{x}_i+\mathbf{PE}_i
\end{aligned}
\end{equation}
\noindent
where $d_e$ is the dimension of the vector representation, $i$ represents the index of the token in the sequence. $\mathbf{PE}_i$ indicates that the sine variable is added to the even position of the token vector whist the cosine variable is added to the odd position. Thereafter, the output token vector $\mathbf{x}_i$ is updated by supplementing $\mathbf{PE}_i$, i.e., $\mathbf{p}_i=\mathbf{x}_i+\mathbf{PE}_i$. By doing so, the relative position between different embedding can be inferred without demanding costs .

% 加不共享参数

%!TeX root=main.tex

\section{End-to-End Testing with \texttt{DeepSQLi}}
\label{sec:approach}

\begin{figure}[t!]
  \centering
  \includegraphics[width=.5\linewidth]{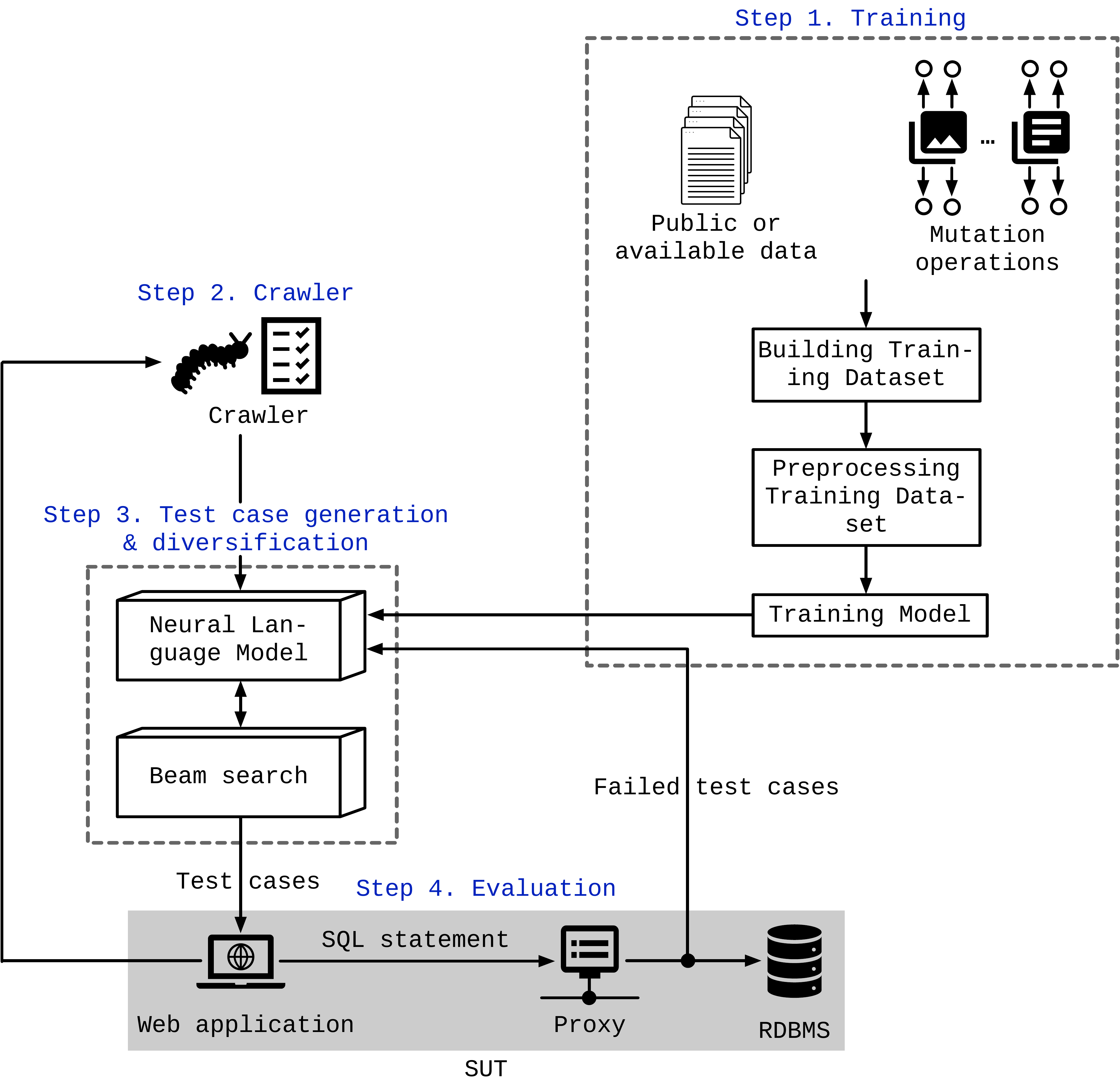}
  \caption{The architecture and workflow of \texttt{DeepSQLi}.}
  \label{fig:tool}
\end{figure}

\texttt{DeepSQLi} is designed as an end-to-end tool, covering all stages in the penetration testing~\cite{HalfondWCSOA09}. As shown in Figure \ref{fig:tool}, the main workflow of \texttt{DeepSQLi} consists of four steps: \textit{training}, \textit{crawler}, \textit{test case generation \& diversification} and \textit{evaluation}, each of which is outlined as follows:
\begin{enumerate}[start=1,label={Step~\arabic*:}]
    \item\textit{\underline{Training:}} This is the first step of \texttt{DeepSQLi} where a neural language model is trained by the \texttt{Transformer}. Agnostic to the Web application, the training dataset can either be summarised from historical testing repository or, as what we have done in this work, mined from publicly available SQLi test cases/data. The collected test cases/data would then be paired, mutated and preprocessed to constitute the training dataset. We will discuss this in detail in \pref{sec:Seq2Seq}.
    
    \item\textit{\underline{Crawler:}} Once the model is trained, \texttt{DeepSQLi} uses a crawler (e.g., the crawler of the Burp Suite project\footnote{~\href{https://portswigger.net/burp}{https://portswigger.net/burp}}) to automatically parse the Web links of the SUT (as provided by software engineers). The major purpose of the crawler is to extract the fields for user inputs in the HTML elements, e.g., \texttt{<input>} and \texttt{<textarea>}, which are regarded as the injection points for a SQLi attack. These injection points, along with their default values, serve as the starting point for the neural language model to generate SQLi test cases. 
    
    \item\textit{\underline{Test Case Generation \& Diversification:}} The neural language model of \texttt{DeepSQLi} is able to generate tokens with different probabilities leading to a test case. To fully exploit such advantage for exploring a diverse set of alternative test cases, in the translation phase, we let the neural language model to generate and explore tokens with the top $m$ probability instead of merely using only the highest one. This is achieved by \texttt{Beam search}~\cite{RaychevVY14}, a heuristic that guides the neural language model to generate $m$ test cases based on the ranked probabilities. This will be elaborated in detail in \pref{sec:diversify}.
    
    \item\textit{\underline{Evaluation:}} Based on the test cases generated at Step 3, we randomly choose one and fed it into the SUT for evaluation. In particular, to avoid compromising the SUT, it is equipped with a proxy, i.e., SQL Parser\footnote{~\href{http://www.sqlparser.com}{http://www.sqlparser.com}.}, before the RDBMS to identify whether or not a malicious SQL statement can achieve a successful attack. To improve the chance of detecting different vulnerabilities, \texttt{DeepSQLi} stops exploring a specific vulnerability once it has been found through a test case.
\end{enumerate}

\vspace{2mm}

It is worth noting that \texttt{DeepSQLi} does not discard unsuccessful test cases which fail to achieve attacks as identified by the proxy. Instead, they are fed back to the neural language model to serve as a starting point for a re-translation (i.e., go through Step 3 again). This comes up with a closed-loop until the test case successfully injects the SUT or the maximum number of attempts is reached. By this means, \texttt{DeepSQLi} grants the ability to generate not only the standard SQLi attacks, but also those more sophisticated ones which would otherwise be difficult to create.

\subsection{Training of Neural Language Model}
\label{sec:Seq2Seq}

\texttt{DeepSQLi} is agnostic to the SUT since we train a neural language model to learn the semantic knowledge of SQLi attack that is independent to an actual Web application. Therefore, \texttt{DeepSQLi}, once being trained sufficiently, can be applied to a wide range of SUT as the semantic knowledge of SQLi is easily generalisable. The overall training procedure is illustrated in~\pref{fig:tool}. In the following subsections, we further elaborate some key procedures in detail.

\subsubsection{Building Training Dataset}
\label{sec:dataset}

In practice, it is possible that the SUT has accumulated a good amount of test cases from previous testing runs, which can serve as the training dataset. Otherwise, since we are only interested to learn the SQL semantics for injections, the neural language model of \texttt{DeepSQLi} can be trained with any publicly available test cases for SQLi testing regardless to the Web applications, as what we have done in this paper.

Since our purpose is to translate and generate a semantically related test case based on either a normal user inputs or another test case, the test cases in the training dataset, which work on the same injection point(s), need to appear in input-output pairs for training. In particular, an input-output pair $(A,B)$ is valid if any of the following conditions is met:
% Specifically, we pair $A$, which may be a set of normal user inputs or a test case, with a test case $B$, where $A$ is the training input and $B$ is the training output, if they conform to any of the following conditions:

\begin{itemize}
    \item $A$ is a known normal user input and $B$ is a test case. For example, \textcolor{blue}{\texttt{https://xxx/id=7}} can be paired with \textcolor{blue}{\texttt{\itshape https://xxx/id=7 union select database()}}.
    
    \item $A$ is a test case whilst $B$ is another one, which is different but still conform to the same type of attack. For example, $A$ is \textcolor{blue}{\texttt{\itshape ' OR 1=1 -- }} and $B$ is \textcolor{blue}{\texttt{\itshape ' OR 5<7 -- }}. It is clear that both of them lead to a semantically related tautology attack.
    
    \item $A$ is a test case whilst $B$ is an extended one based on $A$, thereby we can expect that $B$ is more sophisticated but in a different attack type. For example, $A$: \textcolor{blue}{\texttt{\itshape ' OR 1=1; -- }} can be paired with $B$: \textcolor{blue}{\texttt{\itshape ' or 1=1; select group\_concat(schema\_ name) from information\_schema. schema" \itshape ta; --}}, which belongs to a different type of attack, i.e., a piggybacked queries attack extended from $A$.
\end{itemize}
In this work, we manually create input-output test case pairs to build the training dataset based on publicly available SQLi test cases, such as those from public repositories, according to the aforementioned three conditions. More specifically, the training dataset is built according to the following two steps.
\vspace{2mm}
\begin{table}[t!]
\centering
\tiny
\caption{Description of five mutation operators used in \texttt{DeepSQLi} to enrich the training dataset.}
\label{tab:mutation}
\begin{tabular}{@{}clcc@{}}
\toprule
\multirow{2}{*}{\begin{tabular}[c]{@{}c@{}}Operators\end{tabular}} & \multicolumn{1}{c}{\multirow{2}{*}{Explanation}} & \multicolumn{2}{c}{Example} \\ & \multicolumn{1}{c}{} & Input & Output \\ \cmidrule(r){1-4}
\texttt{Predicate} & \begin{tabular}[c]{@{}l@{}}Mutation by using relational predicates without changing the expression's \\ logical results. In particular, the relational predicates are \{$<,>,\leq,\geq,between,in,like$\}.\end{tabular} & \textcolor{blue}{\texttt{\itshape and 8\textgreater{}= 56}} & \textcolor{blue}{\texttt{\itshape and\lq l\rq in (\lq m\rq,\lq y\rq)}} \\\cmidrule(r){1-4}
\texttt{Unicode} & Mutation from a character to its equivalent Unicode format. & \textcolor{blue}{\texttt{\itshape \#}} & \textcolor{blue}{\texttt{\itshape \%23 }}\\\cmidrule(r){1-4}
\texttt{ASCII} & Mutation from a character to its equivalent ASCII encoding format. & \textcolor{blue}{ \texttt{\itshape a}} & \textcolor{blue}{\texttt{\itshape CHAR(97)}} \\\cmidrule(r){1-4}
\texttt{Keywords} & Confusion of the capital and small letters of keywords in a test case. & \textcolor{blue}{\texttt{\itshape select}} & \textcolor{blue}{\texttt{\itshape seLeCt}} \\\cmidrule(r){1-4}
\texttt{Blank} & Replace the blank character in a test case with an equivalent symbol. & \textcolor{blue}{ \texttt{\itshape or 1}} & \textcolor{blue}{\texttt{\itshape or/**/1}}\\ \bottomrule
\end{tabular}
\end{table}

\begin{enumerate}[start=1,label={Step~\arabic*:}]
    \item We mined the repositories of fuzzing test or brute force tools from various GitHub projects\footnote{~\href{https://tinyurl.com/wh94b8t}{https://tinyurl.com/wh94b8t}}, given that they often host a large amount of test case data in their library and these data are usually arranged according to the types of attacks (along with normal user inputs). This makes us easier to constitute input-output pairs according to the aforementioned conditions. In particular, it is inevitable to devote non-trivial manual efforts to classifying and arranging some more cluttered data.
    \item When analyzing the mined dataset, we found it is difficult to constitute input-output pairs for the disguise attack. For example, a test case containing \textcolor{blue}{\texttt{\itshape ' OR 1=1 -- }} may fail to inject the SUT due to the existence of the standard input validation. Whereas a successful SQLi attack can be formulated by simply change \textcolor{blue}{\texttt{\itshape ' OR 1=1 -- }} to \textcolor{blue}{\texttt{\itshape  \%27\%20OR\%201=1\%20--}}, which is semantically the same but in a different encoding format. This is caused by the rare existence of semantically similar SQLi attacks based on manipulating synonyms and encoding formats from those public repositories. To tackle this issue, five mutation operators, as shown in~\pref{tab:mutation}, are developed to further exploit the test cases obtained from Step 1. By doing so, we can expect a semantically similar test case that finally conforms to a disguise attack. In principle, these mutation operators are able to enrich the test case samples in the training dataset.
\end{enumerate}

\subsubsection{Preprocessing Training Dataset}
\label{sec:preprocessing}

After building the training dataset, we then need to preprocess the data samples by generalisation and word segmentation to eliminate unnecessary noise in the training data. Notably, unlike classic machine learning approaches~\cite{AriuCTG15} that generalise all the words in a data sample, we only generalise the user inputs, the table name and column name to unify tokens \textcolor{blue}{"\texttt{\itshape [normal]}"}, \textcolor{blue}{"\texttt{\itshape [table]}"} and \textcolor{blue}{"\texttt{\itshape [column]}"}. This is because other words and characters, including text-, encoding-, blank characters-, quotes-transforms, are specifically tailored in a SQLi attack, thereby they should not be generalised. For example, considering a test case \textcolor{blue}{"\texttt{\itshape admin'\%20or\%203<7;\texttt{\itshape --}"}} in the training dataset, it is converted into a sequence as \texttt{"['\textcolor{blue}{[normal]}', '\textcolor{blue}{'}', '\textcolor{blue}{\%20}', '\textcolor{blue}{or}', '\textcolor{blue}{\%2}', '\textcolor{blue}{3}', '\textcolor{blue}{<}', '\textcolor{blue}{7}', '\textcolor{blue}{;}',  '\textcolor{blue}{--}']"} after the generalisation.

\subsubsection{Training the Model}
\label{sec:training}

In \texttt{DeepSQLi}, the neural language model is trained under the \texttt{Transformer} architecture. As suggested by Vaswani et al.~\cite{VaswaniSPUJGKP17}, a stochastic optimization algorithm called \texttt{Adam}~\cite{KingmaB14}, with the recommended settings of $\beta_1=0.9$, $\beta_2=0.98$ and $\epsilon=10^{-9}$, is used as the training algorithm.

To prevent overfitting, a 10-fold cross validation is applied in the training process with \texttt{Adam} to optimize the setting of some hyper-parameters, including the number of layers in the encoder and the decoder, the number of hidden layers and neurons in FNN, as well as the number of heads used in the self-attention mechanism of the \texttt{Transformer}. In particular, the following loss function is used in the training process.
\begin{equation}
\label{eq:objective}
\begin{aligned}
L(\mathbf{Y},P(\mathbf{Y}|\mathbf{X}))&=-\log P\left(\mathbf{y}_1, \ldots, \mathbf{y}_p | \mathbf{x}_1, \ldots, \mathbf{x}_s\right)\\
&=-\sum_{t=1}^{p} \log P\left(\mathbf{y}_t | \mathbf{y}_1, \ldots, \mathbf{y}_{t-1},\mathbf{x}_1, \ldots, \mathbf{x}_s\right)
\end{aligned}
\end{equation}
\noindent
The hyper-parameter setting leading to the minimum of the above loss function is chosen to train the model.

\begin{figure}[t!]
  \centering
  \includegraphics[width=.5\linewidth]{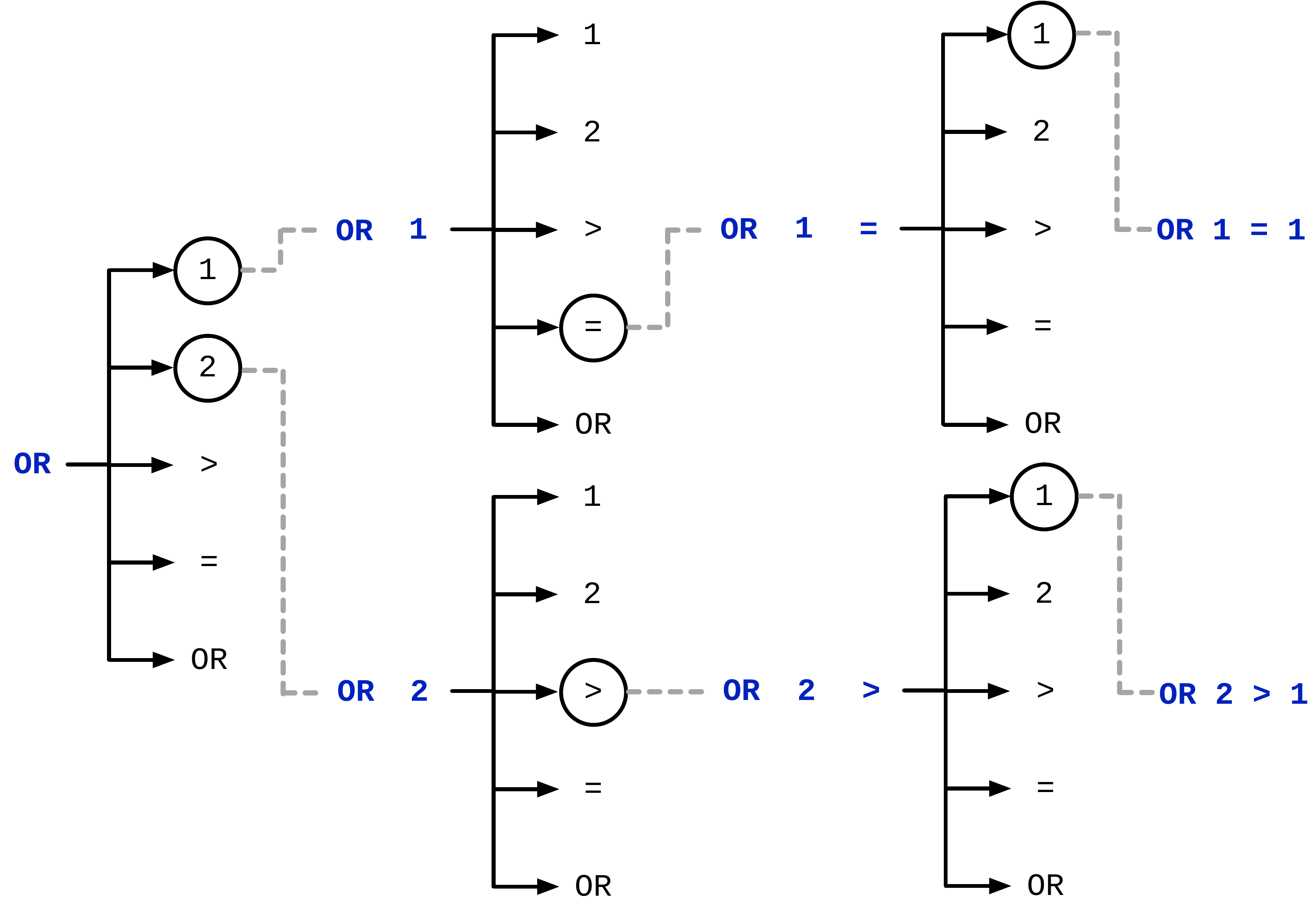}
  \caption{An illustrative example of \texttt{Beam search} with the beam width is 2 and the corpus size is 5, i.e., the possible SQL tokens to be chosen are \textcolor{blue}{\texttt{\itshape "1" }},
  \textcolor{blue}{\texttt{\itshape "2" }},
  \textcolor{blue}{\texttt{\itshape ">" }},
  \textcolor{blue}{\texttt{\itshape "=" }} and
  \textcolor{blue}{\texttt{\itshape "OR" }}
  .}
  \label{fig:beam}
\end{figure}

\subsection{Test Case Generation \& Diversification}
\label{sec:diversify}

Since the goal of the classic machine translation is to identify the most accurate sentence that matches the semantics, only the tokens with the highest probabilities in the context is of interest. In contrast, the major purpose of the test case generation in \texttt{DeepSQLi} is to generate as diverse test cases as possible so that more bugs or vulnerabilities can be identified. In this case, any semantically related test cases are of equal importance as long as they are likely to find new vulnerabilities. By this justification, the classic neural language model, which only outputs the test case having the largest matching probability, is not suitable for \texttt{DeepSQLi}.

To make the generated test cases be more diversified, \texttt{Beam search}~\cite{RaychevVY14} is used to extend and guide the neural language model to generate a set of semantically related test cases. In a nutshell, instead of only focusing on the most probable tokens, \texttt{Beam search} selects the $m$ most probable tokens at each time step given all previously selected tokens, where $m$ is the beam width. Afterwards, by leveraging the neural language model, it merely carries on the search from these $m$ tokens and discard the others. Figure~\ref{fig:beam} shows an example when $m=2$ and the corpus size is 5. According to the first token \textcolor{blue}{" \texttt{\itshape  OR}"} , 2 sequences \textcolor{blue}{ \texttt{\itshape  OR 1}} and \textcolor{blue}{ \texttt{\itshape  OR 2}} with the highest probability are selected from 5 candidate sequences at the first time step. Then, the 2 sequences with the highest probability from the 10 possible output sequences are selected at each subsequent step until the last token in the sequence is predicted. From the above search process, we can also see that \texttt{Beam search} does not only improve the diversity, but also amplify the exploration of search space thus improve the accuracy. In \texttt{DeepSQLi}, we set the beam width as 5, which means that the neural language model creates 5 different test cases for each input. For example, considering the case when the input is \textcolor{blue}{\texttt{\itshape ' OR 1=1}}, then \texttt{DeepSQLi} could generate the following outputs: \textcolor{blue}{\texttt{\itshape ' OR 5<7}}, \textcolor{blue}{\texttt{\itshape ' || 5<7}}, \textcolor{blue}{\texttt{\itshape '+OR+1=1}}, \textcolor{blue}{\texttt{\itshape ' OR 1=1}}, and \textcolor{blue}{\texttt{\itshape ' OR 1=1--}}. 

In principle, this diversification procedure enables \texttt{DeepSQLi} to find more vulnerabilities since the output test cases translated from a given test case (or normal user inputs) are diversified.
% In the \emph{Evaluating} step, we then randomly select one to test the SUT. In this way, \texttt{DeepSQLi} would have larger chance to find more vulnerabilities, as for each given test case (or normal user inputs), the generated one is diversified. 

\section{Evaluation}
\label{sec:experiment}

In this section, the effectiveness of \texttt{DeepSQLi} is evaluated by comparing with \texttt{SQLmap}\footnote{~\href{http://SQLmap.org/}{http://SQLmap.org/}.}, the state-of-the-art SQLi testing automation tool~\cite{Sinha18}, on six real-world Web applications. Note that \texttt{SQLmap} was not designed with automated crawling, thus it is not directly comparable with \texttt{DeepSQLi} under our experimental setup. To make a fair comparison and to mitigate the potential bias, we extend \texttt{SQLmap} with a crawler, i.e., the Burp Suite project used in \texttt{DeepSQLi}.

Our empirical study aims to address the following three research questions (RQs):
\begin{itemize}
  \item \textbf{RQ1:} \textbf{\textit{Is \texttt{DeepSQLi} effective for detecting SQLi vulnerabilities?}}
        
  \item \textbf{RQ2:} \textbf{\textit{Can \texttt{DeepSQLi} outperform \texttt{SQLmap} for detecting SQLi vulnerabilities?}}
  
  \item \textbf{RQ3:} \textbf{\textit{How does \texttt{DeepSQLi} perform on SUT with advanced input validation in contrast to \texttt{SQLmap}?}}
        
\end{itemize}
All experiments were carried out on a desktop with Intel i7-8700 3.20GHz CPU, 32GB memory and 64bit Ubuntu 18.04.2.

\subsection{Experiment Setup}
\label{sec:setup}

\subsubsection{Subject SUT}
\label{sec:sut}

Our experiments were conducted on six SUT\footnote{More detailed information can be found at~\href{http://examples.codecharge.com}{http://examples.codecharge.com}.} written in Java and with MySQL as the back-end database system. All these SUT are real-world commercial Web applications used by many researchers in this literature, e.g.,~\cite{HalfondOM08}. In particular, there are two levels of input validation equipped with these SUT:
\begin{itemize}
    \item \textit{\underline{Essential:}} This level filters the most commonly used keywords in SQLi attacks, e.g., \textcolor{blue}{\lq\texttt{\itshape AND}\rq} and \textcolor{blue}{\lq\texttt{\itshape OR}\rq}.
    \item \textit{\underline{Advanced:}} This is an enhanced level\footnote{This level is usually switched off, because more advanced security mechanism can often make the performance of Web application worse off.} that additionally filters some special characters, which are rarely used but can still be part of a SQLi attack, e.g., \textcolor{blue}{\lq\texttt{\itshape \&\&}\rq} and \textcolor{blue}{\lq\texttt{\itshape ||}\rq}.
\end{itemize}

\pref{tab:webapp} provides a briefing of the SUT considered in our experiments. In particular, these SUT cover a wide spectrum of Web applications under real-world settings. They are chosen from various application domains with different scales in terms of the line-of-code (LOC) and they involve various database interactions (DBIs)\footnote{~The number of SQL statements that can access user inputs.}. Furthermore, the number of Servlets and the number of known SQLi vulnerabilities (dubbed as KV in~\pref{tab:webapp}) are set the same as the existing study~\cite{HalfondOM08}. It is worth noting that both the number of accessible Servlets and their total amount are shown in~\pref{tab:webapp} since not all Servlets are directly accessible.

To mitigate any potentially biased conclusion drawn from the stochastic algorithm, each experiment is repeated 20 times for both \texttt{DeepSQLi} and \texttt{SQLmap} under every SUT.

\subsubsection{Training Dataset}
\label{sec:training_dataset}

In our experiments, the dataset used to train \texttt{DeepSQLi} is constituted by SQLi test cases, consisting of a diverse type of SQLi attacks, collected from various projects in GitHub. We make the training dataset publicly accessible to ensure that our results are reproducible\footnote{~\href{https://tinyurl.com/wh94b8t}{https://tinyurl.com/wh94b8t}}. Afterwards, we preprocess the training dataset by pairing and mutation according to the steps and conditions discussed in~\pref{sec:dataset}. In particular, the number of paired SQLi test case instances is 19,220, all of which can be directly used for training \texttt{DeepSQLi}. After using the mutation operators, the training dataset is significantly diversified and the number of training data instances is increased to 56,841.

\begin{table}[t!]
\caption{Real-world SUT used in our experiments.}
\label{tab:webapp}
\centering
\begin{tabular}{ccccc}
\toprule
SUT & LOC & Servlets$^\ast$ & DBIs & KV \\ \midrule
\textit{Employee} & 5,658  & 7 (10)   & 23   & 25 \\ \midrule
\textit{Classifieds}      & 10,949 & 6 (14)   & 34   & 18 \\ \midrule
\textit{Portal}            & 16,453 & 3 (28)   & 67   & 39 \\ \midrule
\textit{Office Talk}      & 4,543  & 7 (64)   & 40   & 14 \\ \midrule
\textit{Events}             & 7,575  & 7 (13)   & 31   & 26 \\ \midrule
\textit{Checkers}          & 5,421 & 18 (61)    & 5   & 44 \\ \bottomrule
\end{tabular}
\begin{tablenotes}
    \item[1] $^\ast$ The $\#$ of accessible Servlets (the total $\#$ of Servlets).
\end{tablenotes}
\end{table}

\subsubsection{Quality Metrics}
\label{sec:metrics}

The following three quality metrics are used in our empirical evaluations.

\begin{itemize}
    \item \textbf{Number of vulnerabilities found:} We use the number of SQLi vulnerabilities identified by either \texttt{DeepSQLi} or \texttt{SQLmap} as a criterion to evaluate their ability for improving the security of the underlying SUT.

    \item \textbf{Number of test cases and exploitation rate (ER):} In order to evaluate the ability for utilising computational resources, we keep a record of the total number of test cases generated by either \texttt{DeepSQLi} or \texttt{SQLmap}, denoted as $T_{\texttt{total}}$. In addition, we also chase the number of test cases that successfully lead to SQLi attacks, denoted as $T_{\texttt{success}}$. Thereafter, ER is the ratio of $T_{\texttt{success}}$ to  $T_{\texttt{total}}$, i.e., $\frac{T_{\texttt{success}}}{T_{\texttt{total}}}$.

    \item \textbf{CPU wall clock time:} In order to evaluate the computational cost required by either \texttt{DeepSQLi} or \texttt{SQLmap}, we keep a record of the CPU wall clock time used by them for testing the underlying SUT.
\end{itemize}

\vspace{2mm}
\begin{figure}[t!]
  \centering
  \includegraphics[width=.5\linewidth]{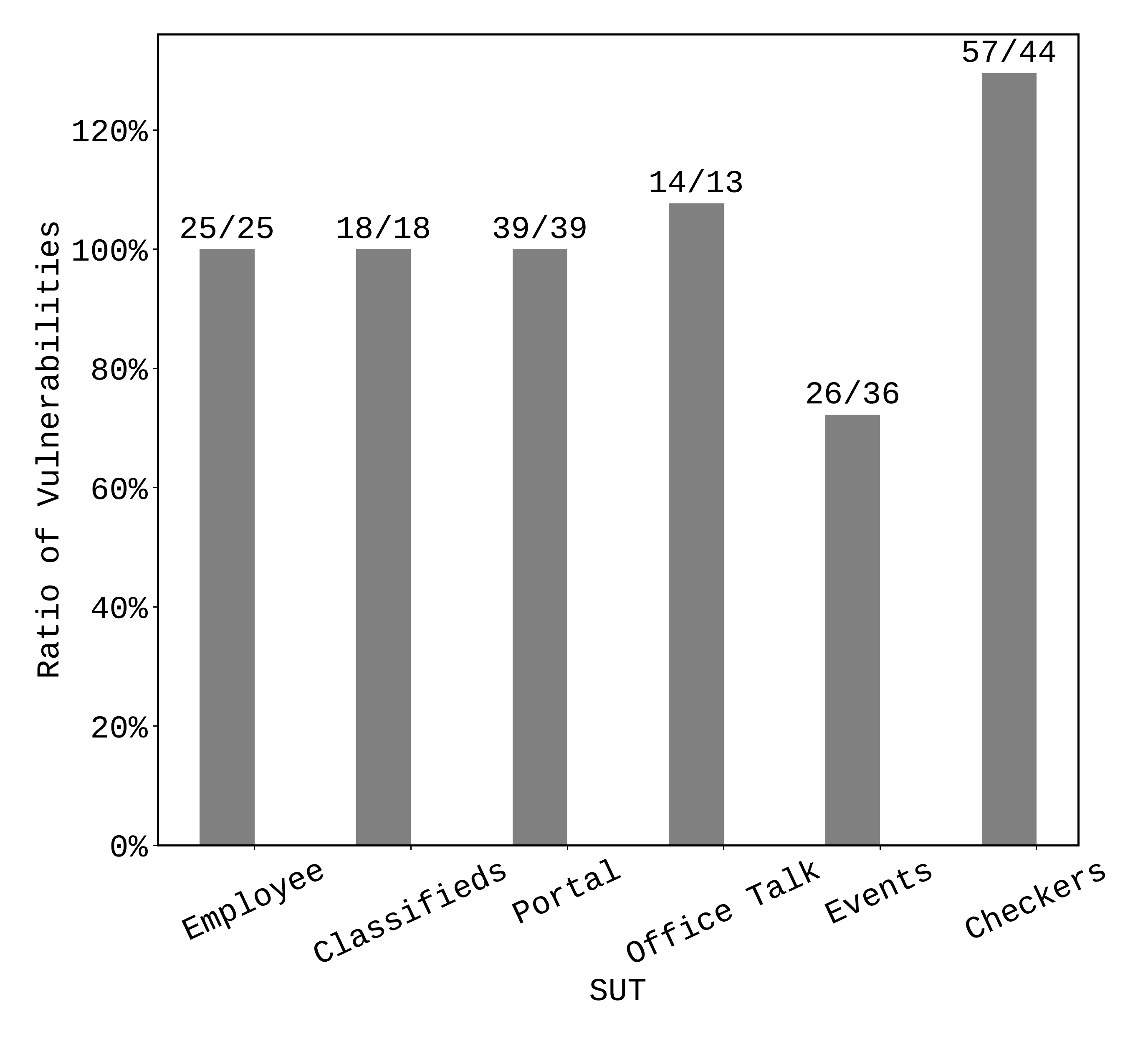}
  \caption{Ratio of $\#$ of vulnerabilities identified by \texttt{DeepSQLi} to that of \texttt{SQLmap}.}
  \label{fig:rq1}
\end{figure}

\subsection{The Effectiveness of \texttt{DeepSQLi}}

\vspace{2mm}
\begin{table}[t!]
% \footnotesize
\caption{Examples of inputs and outputs of \texttt{DeepSQLi}}
\label{tab:example}
\centering
\scalebox{0.7}{
\begin{tabular}{c|c|l}
\hline
\multirow{2}{*}{Example 1} & \textit{Input}  & \texttt{\itshape \textbf{SELECT} card\_type\_id \textbf{FROM} card\_types \textbf{WHERE} card\_type\_name=\lq \ \textcolor{blue}{and+1=(select count(*) from );--} \rq;}                                                                                                            \\ \cline{2-3} 
                   & \textit{Output} & \texttt{\itshape \textbf{SELECT} card\_type\_id \textbf{FROM} card\_types \textbf{WHERE} card\_type\_name=\lq \ \textcolor{blue}{\rq and+1=(select count(*) from ); --} \rq;}                                                                                                              \\ \hline\hline
\multirow{4}{*}{Example 2} & \textit{Input}  & \texttt{\itshape \textbf{SELECT} category\_id, name, par\_category\_id  \textbf{FROM} categories  \textbf{WHERE} category\_id= \textcolor{blue}{and 3=4} ;}                                                                                                                      \\ \cline{2-3} 
                   & \textit{Output} & \texttt{\itshape \textbf{SELECT} category\_id, name, par\_category\_id \textbf{FROM} categories \textbf{WHERE} category\_id= \textcolor{blue}{and \lq\lq7} ;}                                                                                                                      \\ \cline{2-3} 
                   & \textit{Input}  & \texttt{\itshape \textbf{SELECT} category\_id, name, par\_category\_id \textbf{FROM} categories \textbf{WHERE} category\_id= \textcolor{blue}{and \lq\lq9} ;}                                                                                                                       \\ \cline{2-3} 
                   & \textit{Output} & \texttt{\itshape \textbf{SELECT} category\_id, name, par\_category\_id \textbf{FROM} categories  \textbf{WHERE} category\_id= \textcolor{blue}{and\%208=9} ;}                                                                                                                  \\ \hline\hline
\multirow{2}{*}{Example 3} & \textit{Input}  & \texttt{\itshape \textbf{SELECT} category\_id, name, par\_category\_id \textbf{FROM} categories \textbf{WHERE} category\_id= \textcolor{blue}{2};}                                                                                                                           \\ \cline{2-3} 
                   & \textit{Output} & \texttt{\itshape \textbf{SELECT} category\_id, name, par\_category\_id \textbf{FROM} categories \textbf{WHERE} category\_id= \textcolor{blue}{2;delete from members};}                                                                                                         \\ \hline\hline
\multirow{2}{*}{Example 4} & \textit{Input}  & \texttt{\itshape \textbf{SELECT} id, level \textbf{FROM} members \textbf{WHERE} member\_login =\lq \textcolor{blue}{\rq\%20||\%20\lq h \rq=\lq h} \rq \textbf{AND} member\_password=\lq \textcolor{blue}{\rq \#} \rq; }                                                                                              \\ \cline{2-3} 
                   & \textit{Output} & \texttt{\itshape \textbf{SELECT} id, level \textbf{FROM} members \textbf{WHERE} member\_login =\lq \textcolor{blue}{\rq \%20Or\%20\lq h\rq=\lq h} \rq \textbf{AND} member\_password=\lq \textcolor{blue}{\rq --} \rq; }                                                                                               \\ \hline\hline
\multirow{2}{*}{Example 5} & \textit{Input}  & \begin{tabular}[c]{@{}l@{}}\texttt{\itshape \textbf{INSERT INTO} members (member\_login,member\_password,name,email,location,work\_phone,home\_phone)} \\\texttt{\itshape \textbf{VALUES} ("\textcolor{blue}{select database()}",""\textcolor{blue}{test}",""\textcolor{blue}{test}",""\textcolor{blue}{test}",""\textcolor{blue}{test}",""\textcolor{blue}{1}","\textcolor{blue}{1}")};\end{tabular} \\ \cline{2-3} 
                   & \textit{Output} & \begin{tabular}[c]{@{}l@{}}\texttt{\itshape \textbf{INSERT INTO} members (member\_login,member\_password,name,email,location,work\_phone,home\_phone)} \\\texttt{\itshape \textbf{VALUES} ("\textcolor{blue}{sElEct\**\ database()}","\textcolor{blue}{test}","\textcolor{blue}{test}","\textcolor{blue}{test}","\textcolor{blue}{test}","\textcolor{blue}{1}","\textcolor{blue}{1}")};\end{tabular} \\ \hline
\end{tabular}}
\end{table}

In this section, we firstly examine \texttt{DeepSQLi} on SUT with the \textit{essential} input validation. \pref{fig:rq1} presents the ratio of the average number of vulnerabilities identified by \texttt{DeepSQLi} (over 20 runs) to that of known vulnerabilities. From this result, we can clearly see that \texttt{DeepSQLi} is able to identify all known SQLi vulnerabilities for 5 out 6 SUT, except the \textit{Events}. This might be caused by the crawler which fails to capture all injection points in \textit{Events}. In contrast, it is worth noting that \texttt{DeepSQLi} is able to identify more SQLi vulnerabilities than those reported in~\cite{HalfondO05} for \textit{Office Talk} and \textit{Checker}. This is a remarkable result that demonstrates the ability of \texttt{DeepSQLi} for identifying previously unknown and deeply hidden vulnerabilities of the underlying black-box SUT.

To further understand why \texttt{DeepSQLi} is effective for revealing the SQLi vulnerabilities, Table~\ref{tab:example} shows the injectable SQL statements and the related test cases generated in our experiments. Specifically, in Example 1, \texttt{DeepSQLi} is able to learn that the input test case, which is an unsuccessful SQLi attack, has failed due to a missing quotation mark. Thereby, it generates another semantically related test case which did find a vulnerability. In Example 2, we see more semantically sophisticated amendments though exploiting the learned semantic knowledge of SQL: the initially failed test case \textcolor{blue}{\texttt{\itshape and 3=4}} is translated into another semantically related one, i.e., \textcolor{blue}{\texttt{\itshape and\lq\lq9}}, which failed to achieve an attack again. Subsequently, in the next round, \texttt{DeepSQLi} then translates it into a more sophisticated and successful test case \textcolor{blue}{\texttt{\itshape and\%208=9}}, which eventually leads to the discovery of a vulnerability. Likewise, for Examples 3 to 5, the input test cases have been translated into another semantically related and more sophisticated test cases.

\vspace{0.5em}
\noindent
\framebox{\parbox{\dimexpr\linewidth-2\fboxsep-2\fboxrule}{
		\underline{\textbf{Answer to RQ1}}: \textbf{\textit{Because of the semantic knowledge learned from previous SQLi attacks, \texttt{DeepSQLi} has shown its effectiveness in detecting SQLi vulnerabilities. It is worth noting that \texttt{DeepSQLi} is able to uncover more vulnerabilities that are deeply hidden and previously unknown in the SUT.}}
}}

\vspace{-0.5em}
\subsection{Performance Comparison Between \texttt{DeepSQLi} and \texttt{SQLmap}}

Under the SUT with the \textit{essential} input validation, \pref{tab:test-cases} shows the comparison results of the total number of test cases generated by \texttt{DeepSQLi} and \texttt{SQLmap} (dubbed as \#total) versus the amount of test cases leading to successful attacks (dubbed as \#success). From these results, it is clear that \texttt{DeepSQLi} is able to fully test the SUT with fewer test cases than \texttt{SQLmap}. In addition, as demonstrated by the better ER values achieved by \texttt{DeepSQLi}, we can conclude that \texttt{DeepSQLi} is able to better utilise test resources.

\begin{table}[t!]
\centering
\caption{Comparison results of the $\#$ of total/successful test cases generated by \texttt{DeepSQLi} and \texttt{SQLmap}, and the ER values.}
\label{tab:test-cases}
\begin{tabular}{@{}ccccc@{}}
\toprule
\multirow{2}{*}{SUT} & \multicolumn{2}{c}{\texttt{DeepSQLi}} & \multicolumn{2}{c}{\texttt{SQLmap}} \\  
                     & \#total/\#success  & ER      & \#total/\#success & ER     \\ \cmidrule(r){1-5}
\textit{Employee}             & 5563/473           & 8.50\%  & 34851/1534        & 4.40\% \\
\textit{Classifieds}          & 4512/340           & 7.54\%  & 25954/1046        & 4.03\% \\
\textit{Portal}               & 8657/740           & 8.55\%  & 63001/2244        & 3.56\% \\
\textit{Office Talk}          & 2998/260           & 8.67\%  & 9451/462          & 4.89\% \\
\textit{Events}               & 5331/487           & 9.14\%  & 32136/1440        & 4.48\% \\
\textit{Checkers}               & 13463/1077           & 8.00\%  & 67919/3352        & 4.94\% \\
\bottomrule
\end{tabular}
\end{table}

\vspace{2mm}
\begin{figure}[t!]
    \centering
    \includegraphics[width=.5\linewidth]{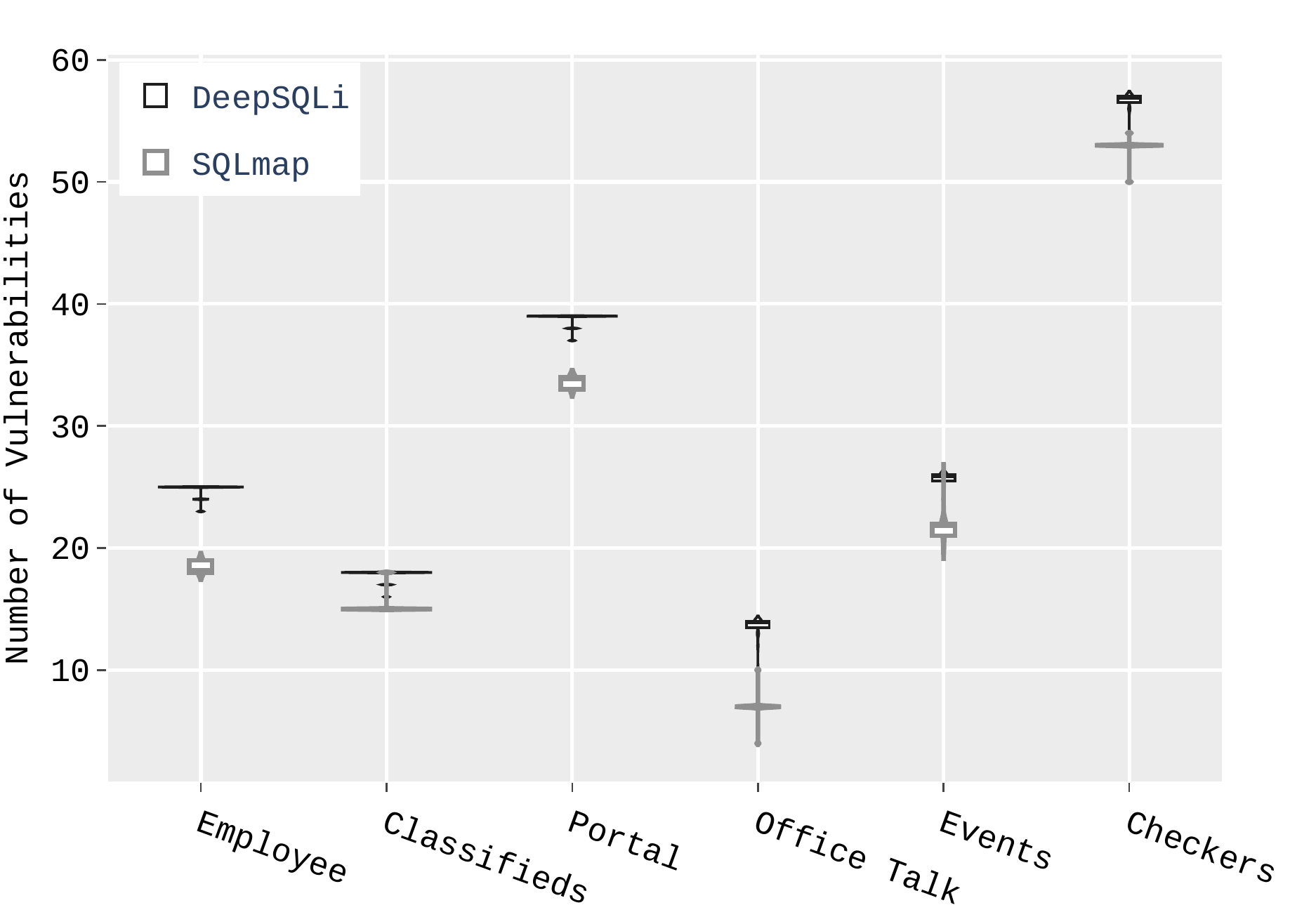}
    \caption{Violin charts of the number of SQLi vulnerabilities identified by \texttt{DeepSQLi} (black lines) and \texttt{SQLmap} (\textcolor{gray}{gray lines}) on six SUT with \textit{essential} input validation across 20 runs.}
    \label{fig:rq2}
\end{figure}

To have an in-depth analysis, \pref{fig:rq2} uses violin charts to visualise the distribution of the number of SQLi vulnerabilities identified by \texttt{DeepSQLi} and \texttt{SQLmap} on all six SUT across 20 runs. From this comparison result, it is clear that \texttt{DeepSQLi} is able to find more vulnerabilities than \texttt{SQLmap} at all instances. In particular, as shown in~\pref{fig:rq2}, the violin charts of \texttt{DeepSQLi} experienced much less variance than that of \texttt{SQLmap}. This observation implies that it is capable of producing more robust results by learning and leveraging the semantic knowledge embedded in the previous SQLi test cases.

\pref{tab:time} shows the comparison results of the CPU wall clock time required for running \texttt{DeepSQLi} and \texttt{SQLmap}. From this comparison result, we find that \texttt{DeepSQLi} runs much faster than \texttt{SQLmap}. In particular, it achieves up to 6$\times$ faster running time at the SUT \textit{Portal}. By cross referencing with the results shown in~\pref{tab:test-cases}, we can see that \texttt{SQLmap} generates much more test cases than \texttt{DeepSQLi}. It is worth noting that more test cases do not indicate any better contribution for revealing SQLi vulnerabilities, because those test cases might be either unsuccessful or redundant as reflected by the lower ER values achieved by \texttt{SQLmap}. In addition, generating a much higher number of (useless) test cases makes \texttt{SQLmap} much slower than \texttt{DeepSQLi}.

\vspace{0.5em}
\noindent
\framebox{\parbox{\dimexpr\linewidth-2\fboxsep-2\fboxrule}{
		\underline{\textbf{Answer to RQ2}}: \textbf{\textit{\texttt{DeepSQLi} is able to find significantly more SQLi vulnerabilities than  \texttt{SQLmap}, with a better utilization of the testing resource as evidenced by the better exploitation rates. \texttt{DeepSQLi} also runs much faster than \texttt{SQLmap} on up to 6$\times$ better.}}
}}

\begin{table}[t!]
\centering
\caption{Comparison of the CPU wall clock time (in second) used to run \texttt{DeepSQLi} and \texttt{SQLmap} over all 20 runs.}
\label{tab:time}
\begin{tabular}{ccc}
\toprule
% \multirow{2}{*}{SUT} & \multicolumn{2}{c}{Runing Time (Seconds)} \\ 
SUT                     & \texttt{DeepSQLi}             & \texttt{SQLmap}             \\ \cmidrule(r){1-3}
\textit{Employee}             & 355                  & 1177               \\
\textit{Classifieds}          & 236                  & 931                \\
\textit{Portal}               & 357                  & 2105               \\
\textit{Office Talk}          & 166                  & 384                \\
\textit{Events}               & 259                  & 1094               \\
\textit{Checkers}             & 519                  & 2228               \\\bottomrule
\end{tabular}
\end{table}

\subsection{Performance Comparison Between \texttt{DeepSQLi} and \texttt{SQLmap} on SUT with Advanced Input Validation}

The previous subsections have validated the effectiveness and performance of \texttt{DeepSQLi} on SUT with \textit{essential} input validation. In this subsection, we switch on the \textit{advanced} input validation in the SUT, aiming to assess the performance of \texttt{DeepSQLi} against \texttt{SQLmap} under more complicated and challenging scenarios. 

\begin{figure}[t!]
  \centering
  \includegraphics[width=.5\linewidth]{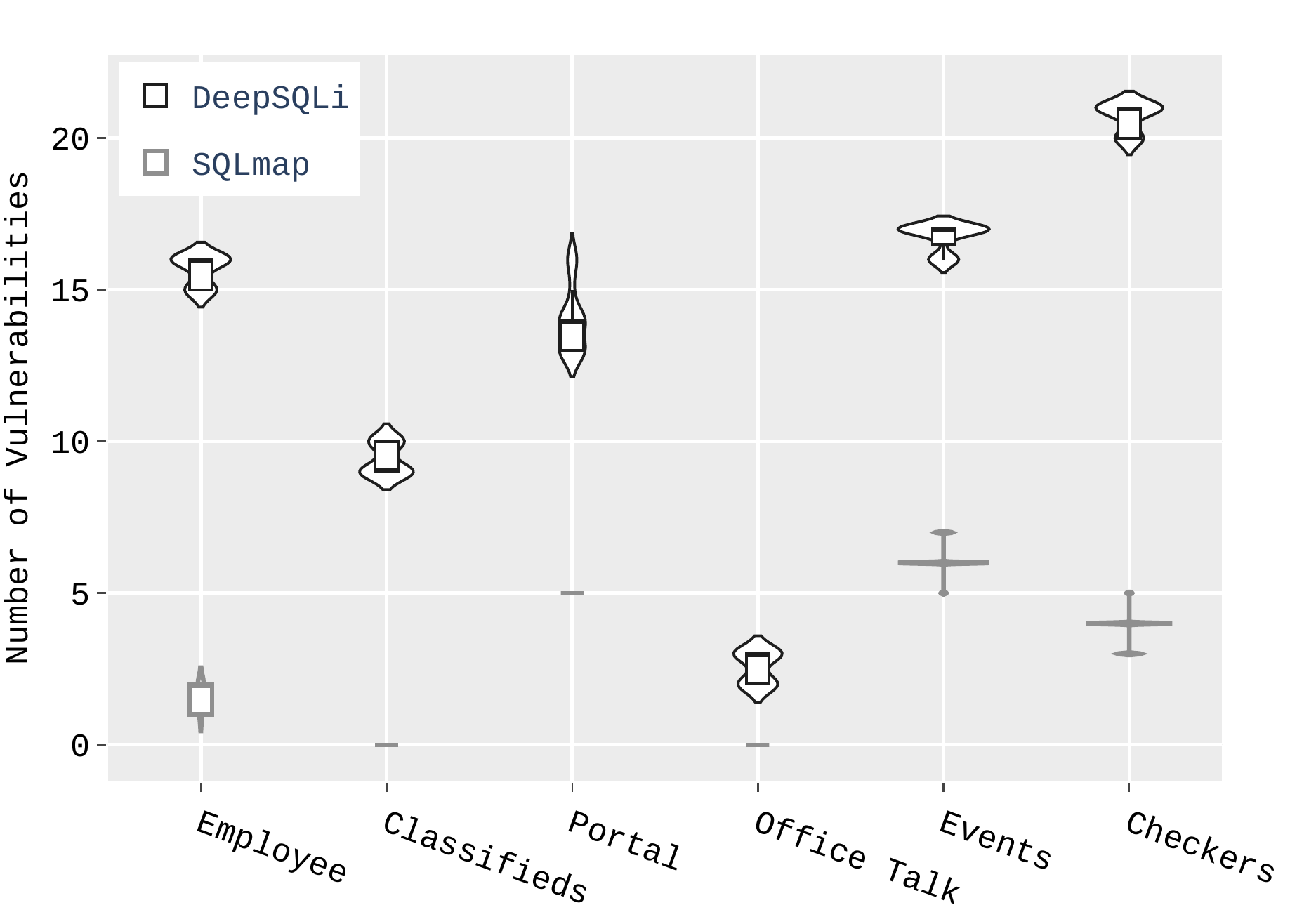}
  \caption{Violin charts of the number of SQLi vulnerabilities identified by \texttt{DeepSQLi} (black lines) and \texttt{SQLmap} (\textcolor{gray}{gray lines}) on SUT with \textit{advanced} input validation across 20 runs.}
  \label{fig:rq3-1}
\end{figure}

\pref{fig:rq3-1} presents the comparison results for the number of SQLi vulnerabilities found by \texttt{DeepSQLi} and \texttt{SQLmap} on SUT under \textit{advanced} input validation. As can be seen, \texttt{DeepSQLi} finds remarkably more SQLi vulnerabilities than \texttt{SQLmap}. It is worth noting that there are a few runs where \texttt{SQLmap} failed to detect any vulnerability at \textit{Employee}, \textit{Classifieds} and \textit{Office Talk}. In contrast, \texttt{DeepSQLi} shows consistently better performance for finding considerably more SQLi vulnerabilities.

\vspace{2mm}
\begin{figure}[t!]
  \centering
  \includegraphics[width=.5\linewidth]{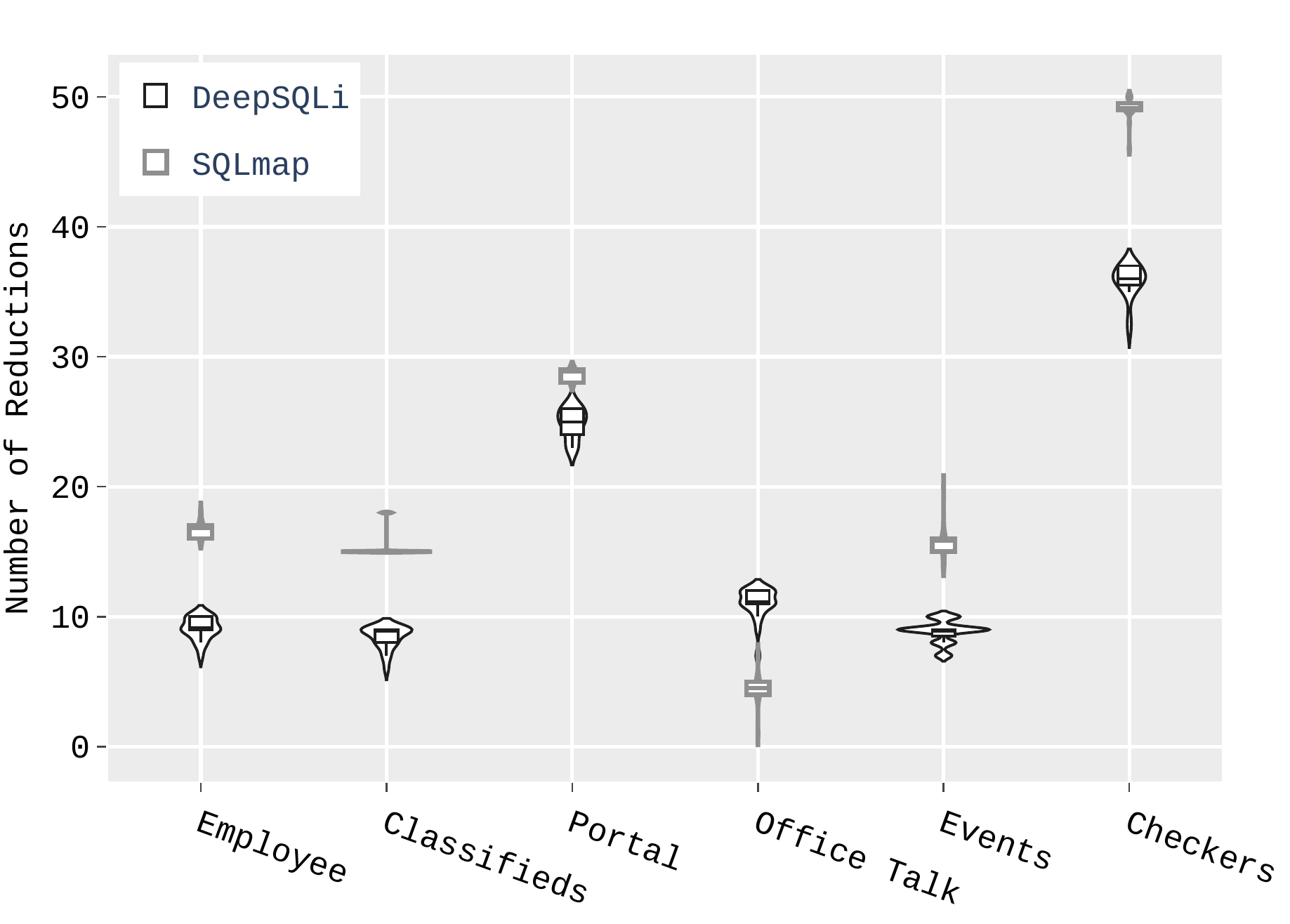}
  \caption{Violin charts of the number of reductions of SQLi vulnerabilities identified by \texttt{DeepSQLi} (black lines) and \texttt{SQLmap} (\textcolor{gray}{gray lines}) on SUT with \textit{advanced} input validation across 20 runs.}
  \label{fig:rq3-2}
\end{figure}

In order to evaluate the sensitivity of \texttt{DeepSQLi} and \texttt{SQLmap} to the strength of input validation, we compare the number of vulnerabilities found on SUT under \textit{advanced} input validation with that on SUT under \textit{essential} input validation. By cross referencing~\pref{fig:rq1}, we can observe some reductions on the number of vulnerabilities identified by both tools, as shown in \pref{fig:rq3-2}. However, in contrast to \texttt{SQLmap}, it is clear that \texttt{DeepSQLi} is much less affected by the strengthened input validation in 5 out of 6 SUT, demonstrating its superior capability of revealing SQLi vulnerabilities in more complicated scenarios. This better result achieved by \texttt{DeepSQLi} can be attributed to the effective exploitation of the semantic knowledge learned from previous SQLi test cases~\cite{LiZZL09,LiZLZL09,LiKWCR12,LiKCLZS12,CaoKWL12,LiKWTM13,LiK14,CaoKWL14,WuKZLWL15,LiKZD15,LiKD15,LiDZ15,LiODY16,LiDZZ17,WuKJLZ17,WuLKZZ17,LiWKC13,ChenLY18,KumarBCLB18,WuLKZ20,ChenLBY18,LiCFY19,WuLKZZ19,LiuLC19,LiXT19,ZouJYZZL19,BillingsleyLMMG19,GaoNL19,LiDY18,Li19,LiCSY19}.

\vspace{0.5em}
\noindent
\framebox{\parbox{\dimexpr\linewidth-2\fboxsep-2\fboxrule}{
		\underline{\textbf{Answer to RQ3}}: \textbf{\textit{Under the \textit{advanced} input validation, \texttt{DeepSQLi} leads to much better results than that of \texttt{SQLmap}, which can hardly find any vulnerability at all in a considerable number of runs. In general, \texttt{DeepSQLi} is much less affected by the Web applications with strengthened input validation.}}
}}

\section{Threats to Validity}
\label{sec:threats}

As with any empirical study, the biases from the experiments can affect the conclusion drawn. As a result, we study and conclude this work with the following threats to validity in mind.
%We have identified the a few threats to validity as below:

The metrics and evaluation method used is a typical example of the construct threats, which concern whether the metrics/evaluation method can reflect what we intend to measure. In our work, the metrics studied are widely used in this field of research~\cite{ThomeGZ14, BozicGSW15} and they serves as quality indicator for different aspects of SQLi testing. To mitigate the randomness introduced by the training, we repeat 20 experiment runs for each tool under a SUT. To thoroughly report our results without losing information, the distributions about the number of SQLi vulnerability found, which is the most important metric, have also been plotted in violin charts.

Internal threats are concerned with the degree of control on the studies, particularly related to the settings of the deep learning algorithm. In our work, the hyperparameters of \texttt{Transformer} are automatically tuned by using \texttt{Adam} and 10-fold cross validation, which is part of the training. The internal parameters of \texttt{Adam} itself were configured according to the suggestions from Vaswani et al.~\cite{VaswaniSPUJGKP17}. The crawler and proxy \texttt{DeepSQLi} are also selected based on their popularity, usefulness and simplicity.

External threats can be linked to the generalisation of our findings. To mitigate such, firstly, we compare \texttt{DeepSQLi} with a state-of-the-art tool, i.e., \texttt{SQLmap}. This is because \texttt{SQLmap} is the most cost-effective tool and has been widely used as a baseline benchmark~\cite{ThomeGZ14, AppeltNPB18, AppeltAB13}. Secondly, we study six real-world SUT that are widely used as standard benchmarks for SQLi testing research~\cite{HalfondOM08, HalfondO05, HalfondOM06}. Despite that all the SUT are based on Java, they come with different scales, number of vulnerabilities and the characteristics, thus they are representatives of a wide spectrum of Web applications. In future work, we aim to evaluate \texttt{DeepSQLi} on other Web applications developed in different programming languages.

%!TeX root=main.tex

\vspace{-1.5em}
\section{Related Work}
\label{sec:survey}
In order to detect and prevent SQLi attacks, different approaches have been proposed over the last decade, including anomalous SQL query matching, static analysis, dynamic analysis, \emph{etc}.
 
Several approaches aim to parse or generate SQLi statements based on specific SQL syntax. For example, Halfond and Orso~\cite{HalfondO05} proposed AMNESIA, a combination of dynamic and static analysis approach. At the static analysis stage, models of the legitimate queries that applications can generate is automatically built. In the dynamic analysis phase, AMNESIA uses runtime monitoring to check whether dynamically generated queries match the model. Mao et al.~\cite{ChenyuMFG16} presented an intention-oriented detection approach that converted SQL statement into a deterministic finite automaton and detect SQL statement to determine if the statement contains an attack. These aforementioned approaches, unlike \texttt{DeepSQLi}, heavily rely on fixed syntax or source code to estimate unknown attacks. In addition, they are not capable of learning the semantic knowledge from the SQL syntax.

% Thome et al.~\cite{ThomeGZ14} proposed \texttt{BIOFUZZ}, a search-based testing approach to detect SQLi vulnerabilities in web applications. They use fitness functions based on multiple components to measure the likelihood of the SQLi attacks. By constructing test cases through specific context-free grammars and optimized them by the fitness function, this approach exposes vulnerabilities on web application quickly.
%\citeauthor{AppeltNBA14}~\cite{AppeltNBA14} proposed \texttt{$\mu$4SQLi}, an automated testing approach to use mutation operators producing malicious SQL statements. The test cases are all generated by the specific mutation operators developed by them. These mutated test cases increase the likelihood of triggering vulnerabilities and perform well in experiments. In this paper, we also use a little bit of mutation operator as shown in ~\pref{tab:mutation}. But we only use it to constitute some input-output pairs in dataset.

Among other tools, \texttt{BIOFUZZ}\cite{ThomeGZ14} and \texttt{$\mu$4SQLi}\cite{AppeltNBA14} are black-box and automated testing tools that bear some similarities to \texttt{DeepSQLi}. However, they have not made the working source code publicly available or the accessible code is severely out-of-date, thus we cannot compare them with \texttt{DeepSQLi} in our experiments. Instead, here we qualitatively compare them in light with the contributions of \texttt{DeepSQLi}:
\begin{itemize}
    \item \texttt{BIOFUZZ}~\cite{ThomeGZ14} is a search-based tool that generates test cases using context-free grammars based fitness function. However, as the fitness function is artificially designed based solely on prior knowledge, it is difficult to fully capture all possible semantic knowledge of SQLi attacks. This is what we seek to overcome with \texttt{DeepSQLi}. Further, the fact that \texttt{BIOFUZZ} relies on fixed grammars may also restrict its ability to generate semantically sophisticated test cases.
    \item \texttt{$\mu$4SQLi}~\cite{AppeltNBA14} is an automated testing tool that uses mutation operators to modify the test cases, with a hope of finding more SQLi vulnerabilities. However, these mutation operators are designed with a set of fixed patterns, thus it is difficult to generate new attacks that have not been captured in the patterns. In \texttt{DeepSQLi}, we also design a few mutation operators, but they are solely used to enrich the training data, which would then be learned by the neural language model. In this way, \texttt{DeepSQLi} is able to create attacks that have not been captured by patterns in the training samples.
\end{itemize}

\texttt{SQLMap} is used as state-of-the-art in the experiments because it is a popular and actively maintained penetration SQLi testing tool, which has been extensively used in both academia~\cite{AppeltAB13,AppeltNPB18,ThomeGZ14} and industry~\cite{Sinha18}. Here we also make a qualitative comparison of differences between 
\texttt{SQLMap} and \texttt{DeepSQLi}.
\begin{itemize}
	\item \texttt{SQLMap} relies on predefined syntax to generate test cases. Such practice, as discussed in the paper, cannot actively learn and search for new SQLi attacks, as the effectiveness entirely depends on the manually crafted rules, which may involve errors or negligence. On the other hand, \texttt{DeepSQLi} learns the semantics from SQL statements and test cases. Such a self-learning process allows it to generalize to previously unforeseen forms of attacks. Our experiments have revealed the superiority of \texttt{DeepSQLi} in detecting the SQLi vulnerabilities.
	\item \texttt{SQLMap} generates new test cases from scratch. \texttt{DeepSQLi}, in contrast, allows intermediately unsuccessful, yet more malicious test cases to be reused as the inputs to generate new one. This enables it to build more sophisticated test cases incrementally and is also one of the reasons that leads to a faster process of \texttt{DeepSQLi} over \texttt{SQLMap}.
\end{itemize}

%for \texttt{BIOFUZZ}, as a search-based algorithm, it needs to measure the quality of test cases through fitness function, while artificially defined fitness functions are difficult to optimize according to the appropriate scenario. For \texttt{$\mu$4SQLi}, it still relies on known attack patterns to generate test cases, which cannot guarantee the integrity of known patterns on the one hand, and cannot create new attack forms on the other hand. 

%\texttt{DeepSQLi} avoids the disadvantages of both types of tools, so it should be in principle better.
Recently, the combination of machine learning and injection-based vulnerability prevention has become popular~\cite{AriuCTG15}\cite{Sheykhkanloo17}\cite{Sheykhkanloo17}\cite{AppeltNPB18}\cite{SkaruzS07}. Among others, Kim et al.~\cite{KimL14} used internal query trees from the log to train a SVM to classify whether an input is malicious. Sheykhkanloo et al.~\cite{Sheykhkanloo17} trained a neural network with vectors which assigned for attacks to classify SQLi attacks. Appelt et al.~\cite{AppeltNPB18} presented ML-Driven, an approach generates test cases with context-free grammars and train a random forest to detect SQLi vulnerability as the software runs. Jaroslaw et al.~\cite{SkaruzS07} applied neural networks to detect SQLi attacks. Their purpose is to build a model that learns the normal input and predicts whether the next input of user is malicious or not. 

%Sheykhkanloo et al.~\cite{Sheykhkanloo17} train a neural network with vectors which assigned for attacks to simply classify SQLi attacks

Unlike \texttt{DeepSQLi}, none of the work aims to generate SQLi test cases for conducting end-to-end testing on the Web application. Moreover, \texttt{DeepSQLi} leverages deep NLP to explicitly learn the semantic knowledge of the SQL for generating the whole sequence of SQLi test case. In particular, \texttt{DeepSQLi} translates the normal user inputs into test cases, which, when fail, would then be re-entered into \texttt{DeepSQLi} to generate more sophisticated SQLi attacks.

%!TeX root=main.tex

\vspace{-.8em}
\section{Conclusion and Future Work}
\label{sec:conclusion}

SQLi attack is one of the most devastating cyber-attacks, the detection of which is of high importance. This paper proposes \texttt{DeepSQLi}, a SQLi vulnerability detection tool that leverages on the semantic knowledge of SQL to generate test cases. In particular, \texttt{DeepSQLi} relies on the \texttt{Transformer} to build a neural language model under the Seq2Seq framework, which can be trained to explicitly learn the semantic knowledge of SQL statements. By comparing \texttt{DeepSQLi} with \texttt{SQLmap} on six real-world SUT, the results demonstrate the effectiveness of \texttt{DeepSQLi} and its remarkable improvement over the state-of-the-art tool, whilst still being effective on Web applications with \textit{advanced} input validation.

In future work, we will extend \texttt{DeepSQLi} with incrementally updated neural language model by using the generated test cases as the testing runs. Moreover, expanding \texttt{DeepSQLi} to handle other vulnerabilities, e.g., Cross-site Scripting, is also within our ongoing research agenda.

%\section*{Author Contributions}
%Li designed and supervised the research. Liu built the system and carried out experiments. Li and Chen interpreted data and wrote the manuscript. 

\section*{Acknowledgment}
Li was supported by UKRI Future Leaders Fellowship (Grant No. MR/S017062/1).

\bibliographystyle{IEEEtran}
\bibliography{IEEEabrv,Bibliography-File}

\end{document}